\documentclass[%
 reprint,
 amsmath,amssymb,
 aps,
 prb,
]{revtex4-2}

\usepackage{amsmath} 
\DeclareMathOperator{\Tr}{Tr} 
\usepackage{upgreek} 
\usepackage{bbm} 
\usepackage{graphicx} 
\usepackage{dcolumn} 
\usepackage{bm} 

\usepackage[colorlinks]{hyperref} 
\hypersetup{
    colorlinks=true,    
    linkcolor=blue,     
    citecolor=blue,     
    urlcolor=blue       
}

\makeatletter
\def\@bibdataout@revtex#1#2#3#4{%
  \immediate\write\@bibdataout{%
    @article{#1,
      author = {#2},
      title = {#3},
      journal = {#4},
      url = {\@bibdata@url},
      doi = {\@bibdata@doi}
    }%
  }%
}
\makeatother

\newcommand{\img}{\mathrm{i}} 

\begin{document}

\title{Cavity-modified  quantum electron transport in\\multi-terminal devices and interferometers}
\author{Dalin Bori\c{c}i}
\affiliation{Universit\'{e} Paris Cit\'e, CNRS, Mat\'{e}riaux et Ph\'{e}nom\`{e}nes Quantiques, 75013 Paris, France}
\author{Geva Arwas}
\affiliation{Universit\'{e} Paris Cit\'e, CNRS, Mat\'{e}riaux et Ph\'{e}nom\`{e}nes Quantiques, 75013 Paris, France}
\author{ Cristiano Ciuti}
\affiliation{Universit\'{e} Paris Cit\'e, CNRS, Mat\'{e}riaux et Ph\'{e}nom\`{e}nes Quantiques, 75013 Paris, France}

\date{\today}

\begin{abstract}
\noindent 
We theoretically investigate transport affected by cavity-mediated electron hopping in multi-terminal quantum Hall bars, quantum point contacts, and Aharonov-Bohm interferometers. Beyond determining conductances and resistances, we analyze spatially resolved current distributions and local density of states. Our study reveals how cavity-mediated inter-edge scattering impacts quantum magnetotransport in finite-size systems and how the cavity-mediated hopping significantly alters electron quantum interference effects.

\end{abstract}

\maketitle

\section{\label{sec:intro}
    Introduction
}

The effects of vacuum field fluctuations in condensed matter systems, previously considered negligible due to their minimal influence in free space, have recently garnered significant attention \cite{GarciaVidal2021, Schlawin2022, Bloch2022}. Advances in the fabrication of electromagnetic cavities, which enable exceptionally strong photon mode confinement \cite{scalari_science_2012,Faist_nanolett2017,ashida2023cavity,HerzigSheinfux2024}, have brought ultra-strong light-matter coupling \cite{FornDaz2019,FriskKockum2019} into the realm of feasibility. This has transformed these previously overlooked interactions into a vibrant research area, as they are now recognized for their profound impact on modifying and controlling quantum phases of matter.

On the theoretical front, models have been developed to explore these effects on superconductivity \cite{Jaksch, Curtis2019, sentef2018cavity, Dmytruk2022,kozin2024cavity}, topology \cite{mendez2020renyi,Ciuti2021,Nguyen2023, perez2023light,Lin2023, mendez2023edge, Dmytruk2022, nguyenSSH, dmytruk2024hybrid,gomez2024high,shaffer2024entanglement, bacciconi2024topological,yang2024emergent,bacciconi2024theory,dag2024engineering}, and transport \cite{hagenMuller,bartolo2018vacuum,hagenmuller2018cavity,Arwas2023,rokaj2023weakened,winter2023fractional,nguyenSSH}. Experimentally, recent attention has been directed towards magneto-transport properties and quantum Hall systems \cite{ParaviciniBagliani2018,Appugliese2022,japanese_QPC, kuroyama2024coherent,PRX_ETH_2024,Enkner}, as well as the influence on the critical temperature of charge density wave transitions \cite{Jarc2023}.

In a recent theoretical study \cite{Arwas2023}, we developed a framework for quantum electron transport influenced by cavity vacuum fields, focusing on a regime where photon degrees of freedom can be adiabatically eliminated. This work introduced an effective single-electron Hamiltonian that incorporates cavity-mediated electron hopping \cite{Ciuti2021}. Within this framework, we used the Landauer-Büttiker formalism \cite{Datta1995} to compute the conductance of one-dimensional systems and two-terminal quantum Hall systems. This emerging field, still in its infancy, holds significant potential for application to a broad range of mesoscopic quantum systems. In this paper, we extend the framework to multi-terminal devices, enabling the investigation of cavity-modified spatially-resolved current fields and local density of states. These advancements provide a deeper understanding of microscopic transport phenomena in complex quantum systems. Specifically, we examine the role of cavity vacuum fields in multi-terminal quantum Hall bars and quantum point contacts, which have been studied experimentally \cite{Appugliese2022,japanese_QPC}, and predict cavity-induced modifications in Aharonov-Bohm interferometers, a system yet to be explored experimentally.

\section{\label{sec:GevaTH}
    Theoretical framework
}

Let $\hat{H}_0$ be a single-electron tight-binding Hamiltonian, whose states $\{ \vert i \rangle \}_{i \in \mathrm{sites}}$, localized on the lattice sites, form an orthonormal basis for the single-electron Hilbert space. Let $t_{ij}$ denote the hopping coupling between sites $i$ and $j$, and let $\hat{d}^{\dagger}_i$ be the fermion creation operator such that $\hat{d}^{\dagger}_i \vert \mathrm{vac} \rangle = \vert i \rangle$, where $\vert \mathrm{vac} \rangle$ is the vacuum state.  

In a single-mode approximation, the cavity field vector potential can be expressed as 
$
\hat{\bm{A}}(\bm{r}) = \bm{A}_{\mathrm{vac}}(\bm{r})(\hat{a}+\hat{a}^{\dagger})$, 
where $\bm{A}_{\mathrm{vac}}(\bm{r})$ is the spatial profile of the mode, and $\hat{a}^{\dagger}$ and $\hat{a}$ are the photonic mode creation and annihilation operators. Light-matter coupling is introduced via the Peierls substitution, where the Peierls phase operator is $
\hat{\phi}_{ij} = (\hat{a} + \hat{a}^{\dagger}) g_{ij}$, 
with 
$
g_{ij} = -\frac{e}{\hbar} \int_{\bm{r}_i}^{\bm{r}_j} d\bm{r} \cdot \bm{A}_{\mathrm{vac}}(\bm{r})
$.

Following \cite{Arwas2023}, expanding the Peierls factors $e^{\img \hat{\phi}_{ij}}$ to first order in $g_{ij}$, we obtain a paramagnetic coupling proportional to $(\hat{a} + \hat{a}^{\dagger})$. This coupling is treated as a perturbation on $\hat{H}_0$ using the intermediate Hamiltonian technique \cite{intermedHam}. The resulting effective Hamiltonian projected onto the zero-photon subspace is  
$
\hat{H}_{\mathrm{eff}} = \hat{H}_0 + \hat{\Gamma}$, with

$$
\hat{\Gamma} = \sum_{\lambda\lambda'} \Gamma_{\lambda\lambda'} \hat{c}^{\dagger}_{\lambda} \hat{c}_{\lambda'} \, .
$$
Here, $\hat{c}^{\dagger}_{\lambda}$ and $\hat{c}_{\lambda}$ are the creation and annihilation operators for the single-particle eigenstates. Using such basis, the bare Hamiltonian can be written in the diagonalized form   
$$
\hat{H}_0 = \sum_{\lambda} \varepsilon_{\lambda} \hat{c}^{\dagger}_{\lambda} \hat{c}_{\lambda},
$$
where 
$
\hat{c}^{\dagger}_{\lambda} \vert \mathrm{vac} \rangle = \vert \phi_{\lambda} \rangle = \sum_i \phi_{\lambda}(i) \vert i \rangle$.  
The cavity-mediated hopping coupling between single-particle eigenstates reads:
\begin{equation}
    \Gamma_{\lambda\lambda'} = - \sum_{\mu} \left( \frac{\mathrm{sgn}\left[ \varepsilon_{\mu} - \min( \varepsilon_{\lambda}, \varepsilon_{\lambda'} ) \right]}{\left\vert \varepsilon_{\mu} - \frac{\varepsilon_{\lambda} + \varepsilon_{\lambda'}}{2} \right\vert + \hbar \omega_{\mathrm{cav}}} \right) h_{\lambda \mu} h_{\mu \lambda'},
\end{equation}
where
$
h_{\alpha \beta} = \sum_{ij} ( - \img g_{ij} t_{ij} ) \phi_{\alpha}^{\star}(i) \phi_{\beta}(j)$.

Note that the $\hat{\Gamma}$ matrix effectively couples electrons injected near the Fermi energy $E_F$ \cite{Arwas2023}, which is what matters for transport in linear-response regime (limit of vanishing applied voltages).

The effective Hamiltonian can be written in the initial tight-binding basis as 
\[
\hat{H}_{\mathrm{eff}} = \sum_{ij} \tilde{t}_{ij} \hat{d}^{\dagger}_i \hat{d}_j,
\]
where $\tilde{t}_{ij}$ represents the cavity-mediated hopping coupling (for $i \neq j$) and on-site energy shift (for $i = j$). In a nearest-neighbor tight-binding lattice, we will have for non-neighboring sites $t_{ij} = 0$, but instead $\tilde{t}_{ij} \neq 0$. This formalism can account for continuous systems in the limit where the lattice spacing is small enough. In what follows, we generalize the framework described in \cite{Arwas2023} by considering multi-terminal configurations, local density of states, and spatially-dependent current densities.


\subsection{Multi-terminal linear-regime transport}\label{sec:LBFormalism}
Throughout this article, 
we shall examine the responsive behavior of devices 
by driving current through the electrodes they are coupled to.
The Landauer-B\"uttiker formalism 
is a powerful tool that relates current
and voltage drops across the leads (electrodes)
through the scatterer's conductance (or equivalently, resistance)
\cite{Landauer, Buttiker_LandauerExt}.
We provide a brief outline of this formalism here, 
as it will be instrumental later 
in discussing driving currents 
and voltage biases interchangeably.

Consider a scattering device connected
to $N_{\mathrm{L}}$ leads. 
In Landauer-B\"{u}ttiker formalism, 
we can obtain the current $I_p$ 
flowing through lead $p$ via 
$ 
    I_p 
    = 
    \sum_{q = 1}^{N_{\mathrm{L}}}
    \mathrm{G}_{pq} (V_p - V_q) \, ,
$
with $(V_p - V_q)$ being the voltage drop
between leads $p$ and $q$
and $\mathrm{G}_{pq}$ the linear conductance matrix,
an intrinsic property of the device. 
Such a matrix is obtained by 
either solving the scattering problem, 
or equivalently by using the non equilibrium 
Green's functions (NEGF) formalism \cite{FisherLee}, 
which we discuss in Sec. \ref{sec:NEGF}.
The conductance matrix 
obeys the following property
$\sum_p \mathrm{G}_{pq} = \sum_k \mathrm{G}_{qk}$ for all $q$, 
essentially ensuring Kirchoff's law $\sum_p I_p = 0$.
Defining 
$
\tilde{\mathcal{G}}_{pq} 
= 
- \mathrm{G}_{pq} + 
\delta_{pq}
\sum_{k=1}^{N_{\mathrm{L}}}
\mathrm{G}_{pk}
$,
the Landauer Büttiker formula can be re-expressed as:
\begin{equation}
    I_p
    =
    \sum_{q = 1}^{N_{\mathrm{L}}}
    \tilde{\mathcal{G}}_{pq}  V_q \, .
\end{equation}
The $\tilde{\mathcal{G}}$ matrix is singular
and its kernel is of dimension one, 
since
$
\sum_p \tilde{\mathcal{G}}_{pq} = \sum_q \tilde{\mathcal{G}}_{pq} = 0
$.
The voltages are defined with respect to an origin 
(ground voltage). 
By setting $V_{N_{\mathrm{L}}} = 0$, 
the matrix equation reduces to
$ 
    I_p 
    = 
    \sum_{q = 1}^{N_{\mathrm{L}} - 1}
    \tilde{\mathcal{G}}_{pq} V_q
$ 
for $p \in \{1,...,N_{\mathrm{L}} - 1\}$. 
Removing the last row and line from $\tilde{\mathcal{G}}$ , 
we obtain the reduced invertible $(N_{\mathrm{L}}-1)-$dimensional square matrix $\mathcal{G}$,
so we can write the voltages in terms of the imposed currents
using the resistance matrix $\mathcal{R} = \mathcal{G}^{-1}$. This can be written as
\begin{equation}\label{eqn:VIR}
    V_q 
    = 
    \sum_{p = 1}^{N_{\mathrm{L}} - 1}
    \mathcal{R}_{qp} I_p \, ,
\end{equation}
for $q \in \{1,...,N_{\mathrm{L}} - 1\}$.

\subsection{Conductance matrix and local density of states}\label{sec:NEGF}
At a given Fermi energy $E_F$ (we are considering the zero temperature limit), 
the linear conductance matrix \cite{Datta1995} is given by
the Caroli formula:
\begin{equation}\label{caroli}
    \mathrm{G}_{pq}(E_F)
    =
    \frac{e^2}{h}
    \Tr\{
        G^{\mathrm{r}}
        \Gamma_{p}
        G^{\mathrm{a}}
        \Gamma_{q}
    \} \, .
\end{equation}
The quantities involved are the following:
the retarded 
$G^{\mathrm{r}}(E) = (E - \hat{H}_{\mathrm{eff}} - \Sigma^{\mathrm{r}})^{-1}$ and advanced 
$G^{\mathrm{a}} = (G^{\mathrm{r}})^{\dagger}$ 
Green's functions of the device coupled to the leads
and the linewidths associated to each lead, namely
$\Gamma_{q}= \img (\Sigma^{\mathrm{r}}_q - \Sigma^{\mathrm{a}}_q)$.
The self-energies due to the individual leads 
are given by 
$
\Sigma^{\mathrm{r}}_q
=
\tau_{q}^{\dagger} g^{\mathrm{r}}_{q} \tau_{q}
$.
Here, 
the $\tau_{q}^{\dagger}$
are the coupling matrices of the scattering device to the leads, 
while $g^{\mathrm{r}}_{q}$ is the retarded Green's function of the isolated infinite lead,
which we do not need to compute entirely 
(see Appendix \ref{apx:leadGF} on how to obtain the lead's self-energy).
The full self-energy is given by the sum
$
\Sigma^{\mathrm{r}} = \sum_q \Sigma^{\mathrm{r}}_q
$,
since the leads are considered independent.
The retarded Green's functions 
contain all the information 
about the spectrum of the system. 
In particular, 
the electron spectral function is given by 
$
A = -\frac{1}{\pi} \Im{ \{ G^{\mathrm{r}} \} }
$. 
From there, 
we can extract the density of states, 
that is the trace of $A$, 
or the local density of states, 
given by 
$
\rho(\bm{r}) 
= 
-\frac{1}{\pi} 
\Im{ 
\{
\langle \bm{r} \vert
G^{\mathrm{r}}
\vert \bm{r} \rangle
\}
}
$
.
The ${}^\mathrm{r}$ superscript stands for retarded,
whilst ${}^\mathrm{a}$ for advanced. 
Both are related to one another 
through hermitian conjugation.

\subsection{Spatially-dependent current densities}
Here, we derive the averaged non-equilibrium current density.
We follow a derivation similar in fashion 
to the one carried out for bond currents in Ref. \cite{Nikolic_bondCurrents}. The main difference is that we derive the current vector for arbitrary long-range hopping and multi-terminal configurations.
With our tight-binding effective Hamiltonian $\hat{H}_{\mathrm{eff}} = \sum_{ij} \tilde{t}_{ij} \hat{d}^{\dagger}_i \hat{d}_j$, we can define a velocity operator 
$\hat{\bm{v}} = \img / \hbar [\hat{H}_{\mathrm{eff}}, \hat{\bm{r}}]$,
with 
$
\hat{\bm{r}}
=
\sum_{j} \bm{r}_j \vert j \rangle \langle j \vert
$. 
The electronic charge density operator is given by
$
\hat{n}(\bm{r}_j)
=
-e/{\mathcal{V}}_{\mathrm{cell}} \vert j \rangle \langle j \vert
$, 
and the current density operator by
$
\hat{\bm{j}}(\bm{r}_k)
= 
\frac{1}{2}
\left[
\hat{n}(\bm{r}_k)
\hat{\bm{v}}
+
\hat{\bm{v}}
\hat{n}(\bm{r}_k)
\right]
$, where ${\mathcal{V}}_{\mathrm{cell}}$ is the volume of the lattice elementary cell.
In second quantization, 
the current operator reads:
\begin{equation}
    \hat{\bm{j}}(\bm{r}_k)
    =
    -\img\frac{ e }{2 \mathcal{V}_{\mathrm{cell}} }
    \sum_j
    (\bm{r}_j - \bm{r}_k) 
    \left(
        \tilde{t}_{kj} \hat{d}^{\dagger}_k \hat{d}_j
        -
        \tilde{t}_{jk} \hat{d}^{\dagger}_j \hat{d}_k
    \right) \, .
\end{equation}
We note in passing, that in the standard case of tight-binding lattice with only near-neighbor hopping, 
where $t = \hbar^2/(2 m_{\star} a^2)$, 
taking the continuum limit 
would yield the well known continuum result
$
    \hat{\bm{j}}(\bm{r})
    =
    -e
    \frac{\hbar}{m_{\star}}
    \Im
    \left\{
        \hat{\psi}^{\dagger}(\bm{r})
        \bm{\nabla} \hat{\psi} (\bm{r})
    \right\}
$, 
with
$\hat{\psi}^{\dagger}(\bm{r})$
being the electron field operator.
For a system in thermal equilibrium, 
it is convenient to introduce the lesser 
Green's functions:
$
    G^{<}_{ij}(t-t')
    =
    \frac{\img}{\hbar}
    \langle
    \hat{c}^{\dagger}_j(t') \hat{c}_i(t)
    \rangle 
    =
    \frac{1}{2 \pi \hbar}
    \int_{\mathbb{R}}
    dE \, 
    e^{-\frac{\img}{\hbar}(t-t')E}
    G^{<}_{ij}(E)
$.
The averaged current operator 
in the steady state can be expressed as:
\begin{align}
    \bm{j}(\bm{r}_k)
    &=
    \frac{e}{\mathcal{V}_{\mathrm{cell}} h}
    \sum_{j} (\bm{r}_j - \bm{r}_k)
    \int_{\mathbb{R}}
    dE 
    \, 
    \Im{ \left\{
    \tilde{t}_{kj}
    G^{<}_{jk}(E)
    \right\} } \, .
\end{align}
We can account for the leads
using the Keldysh equation in its matrix form
$
G^{<}(E) 
= 
G^{\mathrm{r}}(E) 
\Sigma^{<}(E) 
G^{\mathrm{a}}(E)
$
and the relation 
giving the lesser self-energy 
in terms of the linewidths, namely
$
\Sigma^{<}(E)
=
\sum_{q=1}^{N_{\mathrm{L}}}
f(E-\mu_{q})
\Gamma_q(E)
$
\cite{Meir1992, Nikolic_bondCurrents}, 
with 
$f(x) = 1/(e^{\beta x} + 1)$ the Fermi-Dirac distribution.
In the zero temperature limit ($\beta \to +\infty$), 
$f(x) = \Theta(-x)$, the Heaviside step function.
Let us assume the voltage drops among the different leads 
are small with respect to the all relevant spectrum energy gaps 
(i.e. the Green's functions are 
slowly varying over the energy scale related to the voltages), 
which is the case in the linear response regime. In these conditions, 
we can define a Fermi energy as 
$E_F = \sum_q \mu_q / N_{\mathrm{L}}$ 
and voltages as
$e V_q = \mu_q - E_F$.
We split the integral by 
integrating first up to $E_F$
and then for each lead $q$ from $E_F$ to $\mu_q$.
Since the $eV_q$ are considered as vanishing, 
we get:
\begin{align}\label{currentMultiTerm}
    \nonumber
    \bm{j}(\bm{r}_k)
    &\simeq
    \frac{e}{\mathcal{V}_{\mathrm{cell}} h}
    \sum_{j} (\bm{r}_j - \bm{r}_k)
    \int_{-\infty}^{E_F}
    dE \, 
    \Im{ \left\{
    \tilde{t}_{kj}
    G^{<}_{jk}(E)
    \right\} } \\
    &+
    \frac{e^2}{h}
    \sum_{q=1}^{N_{\mathrm{L}}}
     V_q 
    \sum_{j} 
    \frac{\bm{r}_j - \bm{r}_k}{\mathcal{V}_{\mathrm{cell}}}
    \Im{ \{
        \tilde{t}_{kj} 
        \mathcal{C}^{q}_{jk}(E_F) 
    \} } \, ,
\end{align}
where we have defined
the correlation functions due to 
the individual leads as
$
\mathcal{C}^{q}_{jk}
=
\left[ 
G^{\mathrm{r}}
\Gamma_q
G^{\mathrm{a}}
\right]_{jk}
$.
The first line in Eq.
\eqref{currentMultiTerm}
represents the persistent current density (if any), 
while the second one describes 
the current induced by 
voltage drops (or equivalently currents)
across the leads.

\begin{figure}[t!]
    \centering
    \includegraphics[scale=0.42]{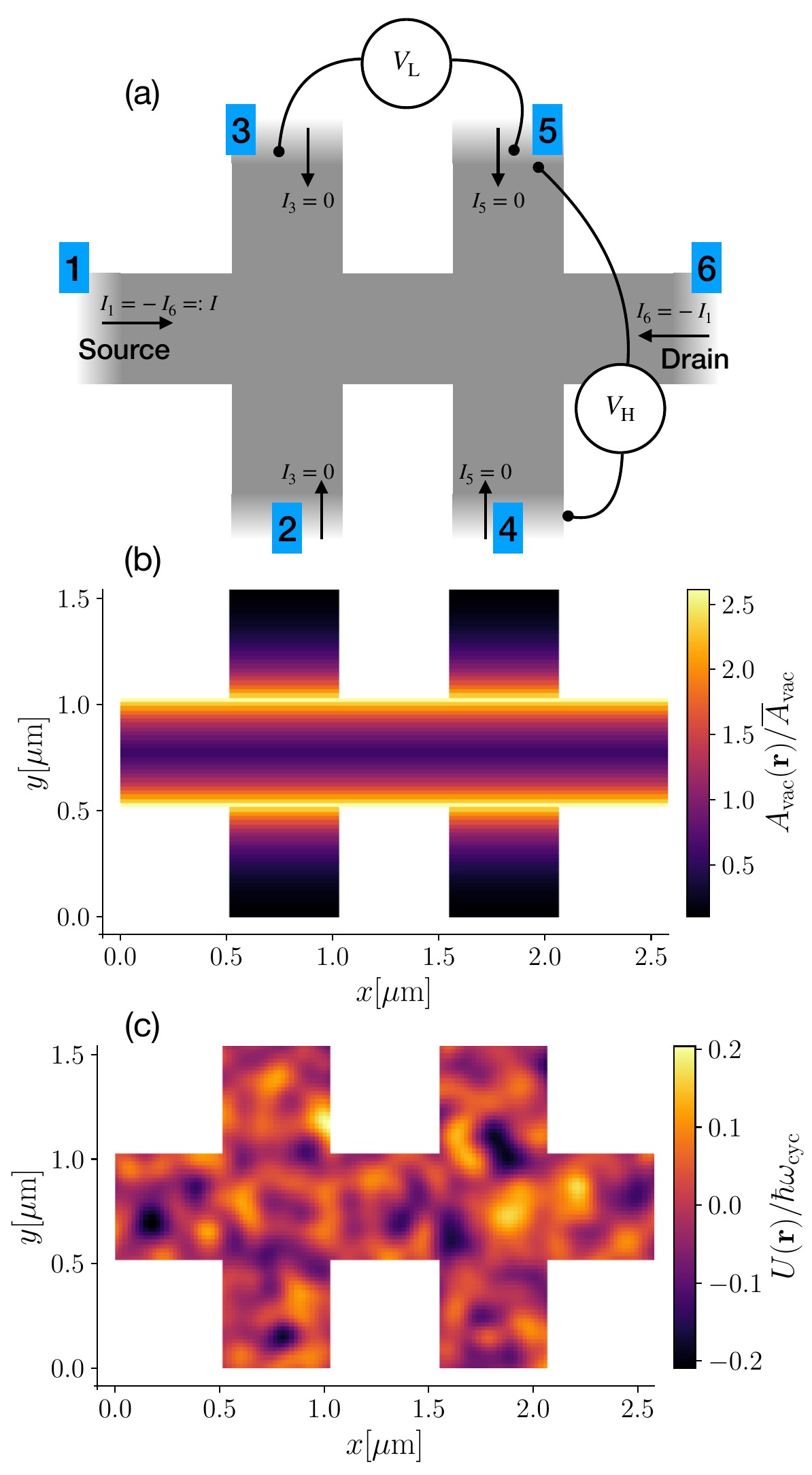}
    \caption{
        (a): sketch of the six-terminal quantum Hall bar. (b): spatial dependence of the considered cavity mode vector potential. (c): disorder energy potential with a finite correlation length, 
        $\lambda_{\mathrm{corr}} = 60 \, \mathrm{nm}$.
        In (a), the dark grey region corresponds to the scattering region, 
        whereas the fading grey ones to the leads. 
        In (b), The field strength is vertically proportional 
        to an exponential decay:
        $A_{\mathrm{vac}}(\bm{r}) \propto e^{-d(y)/\lambda}$, 
        with $d(y)$ being the distance from the closest horizontal edge
        and $\lambda = 0.16 \, \mathrm{\mu m}$, 
        a characteristic length scale. 
    }
    \label{fig:DisProfAndModProf}
\end{figure}

Note that in a tight-binding system with 
only nearest neighbor hoppings, 
the bonds contributing to the current density
are only those connecting nearest neighbor sites. 
In our effective theory, 
the cavity vacuum fields mediate long-range hoppings 
$\tilde{t}_{jk}$, with $j$ and $k$
running throughout the whole sample.

\begin{figure}[t!]
    \centering
    \includegraphics[scale=0.5]{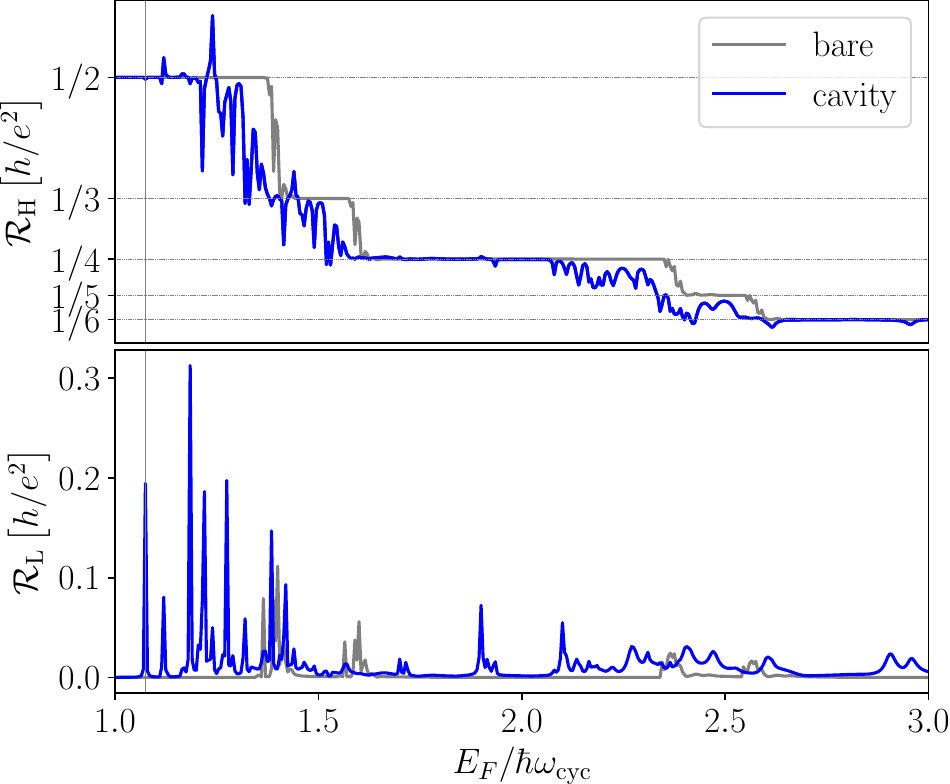}
    \caption{
        Longitudinal (upper panel) and 
        Hall resistances (lower panel)
        of a 2D Hall bar as a function of the Fermi level. 
        The cavity-modified resistance is given in blue, while the bare one (no cavity coupling) in grey.
        The Hall bar is subject to a magnetic field $B= \, 0.1 \mathrm{T}$ 
        and to an electronic disorder
        pictured in Fig. \ref{fig:DisProfAndModProf}.
        Other parameters: discretization grid spacing $a = 10 \, \mathrm{nm}$,  $m_{\star} = 0.067 m_{e}$ (corresponding to GaAs), 
        with $m_{e}$ being the free electron mass, $E_Z = 0.2 \times \hbar \omega_{\mathrm{cyc}}$.
        The cavity field is polarized along the $x$ direction and its amplitude
        is given by 
        $
        A_{\mathrm{vac}}(\bm{r}) 
        = 
        A_{\mathrm{vac}}(y) 
        =
        \Bar{A}_{\mathrm{vac}} 
        \times 
        \mathcal{A}(y)
        $, 
        with 
        $
        \Bar{A}_{\mathrm{vac}} 
        = 
        \frac{\hbar \omega_{\mathrm{cav}}}{2 \pi c e} \sqrt{\alpha_{\mathrm{fsc}} / \eta}
        $ 
        being fixed by the compression factor 
        $\eta = 10^{-10}$ 
        and the cavity mode angular frequency 
        $
        \omega_{\mathrm{cav}} 
        = 
        2 \pi \times 10^{12} \mathrm{rad/s}
        $.
        We include
        $\mathcal{A}(y)$,
        a modulation profile pictured in Fig. \ref{fig:DisProfAndModProf},
        to account for the spatial inhomogeneity of the field. 
        We have introduced $c$ the speed of light 
        and $\alpha_{\mathrm{fsc}} \approx 1/137$ the fine structure constant.
    }
    \label{fig:6term_res}
\end{figure}

\section{Multi-terminal Quantum Hall bar}

As a first example to which we apply our theory, 
we will consider a multi-terminal quantum Hall bar, hosting a
two-dimensional electron gas (2DEG) 
subjected to a perpendicular magnetic field $\bm{B} = B \bm{e}_z$. 
The bare single-electron Hamiltonian is given by:
\begin{equation}\label{spinlessHam}
    \hat{h} = \frac{1}{2m_{\star}} [\hat{\bm{p}} - e \bm{A}_0(\hat{\bm{r}})]^2 + U(\hat{\bm{r}}),
\end{equation}
where 
$
\hat{\bm{p}} 
= 
\hat{p}_x \bm{e}_{x} 
+ 
\hat{p}_y \bm{e}_{y}
$ 
is the electron's momentum, 
$m_{\star}$ is its effective mass, 
and $e$ is the electron charge. 
The operator
$
\hat{\bm{r}} 
= 
\hat{x} \bm{e}_{x} 
+ 
\hat{y} \bm{e}_{y}
$ 
represents the electron's position, 
while $U(\hat{\bm{r}})$ denotes a confining and/or disorder potential. 
The static magnetic field $\bm{B} = \bm{\nabla} \times \bm{A}_0$ can be expressed in terms of a classical vector potential $\bm{A}_0(\hat{\bm{r}})$ .
The presence of the magnetic field influences the electron dynamics 
by quantizing its kinetic energy into discrete, 
equally spaced energy levels known as Landau levels. 
These Landau levels are crucial 
in defining the quantum Hall effect 
and lead to the formation of 
chiral edge modes when the system is finite \cite{girvinBook}.
The energy spacing is directly proportional to the magnetic field $B$ 
and is expressed as 
$E_{\text{cyc}} = \hbar \omega_{\text{cyc}} = \hbar eB / m_{\star}$. 
In order to perform numerical simulations, 
we have discretized this Hamiltonian onto a two-dimensional 
square lattice of spacing $a$. 
The lattice sites are labeled by their positions $\bm{r}_i = (x_i, y_i)$, 
yielding the tight-binding Hamiltonian:
\begin{equation}\label{eqn:Ham2DEG}
    \hat{H} 
    = 
    \sum_i (4t + U_i) \hat{d}_i^\dagger \hat{d}_i - t \sum_{\langle i,j \rangle}  e^{i \phi_{ij}} \hat{d}_i^\dagger \hat{d}_j \, ,
\end{equation}
where $t = \hbar^2 / (2m_{\star} a^2)$ 
is the hopping parameter, 
and $\phi_{ij}$ represents the classical Peierls phase 
due to the static magnetic field. 
In the Landau gauge, 
this phase factor is given by 
$
\phi_{ij} 
= 
y_j a \ell_{\mathrm{cyc}}^{-2}
\delta_{y_i, y_j}
$, 
with $\ell_{\text{cyc}} = \sqrt{\hbar/eB}$ being the magnetic length. 
The potential \( U_i = U(\bm{r}_i) \) 
accounts for any disorder or confinement effects.
The continuum limit is recovered by letting the lattice spacing going to zero.
In all the numerical simulations, we have carefully verified such convergence.
Knowing the bare tight-binding Hamiltonian $\hat{H}$, we have calculated the cavity-mediated effective Hamiltonian $\hat{H}_{\mathrm{eff}}$ with the procedure described above.

\begin{figure*}[t!]
    \centering
    \includegraphics[scale=0.8]{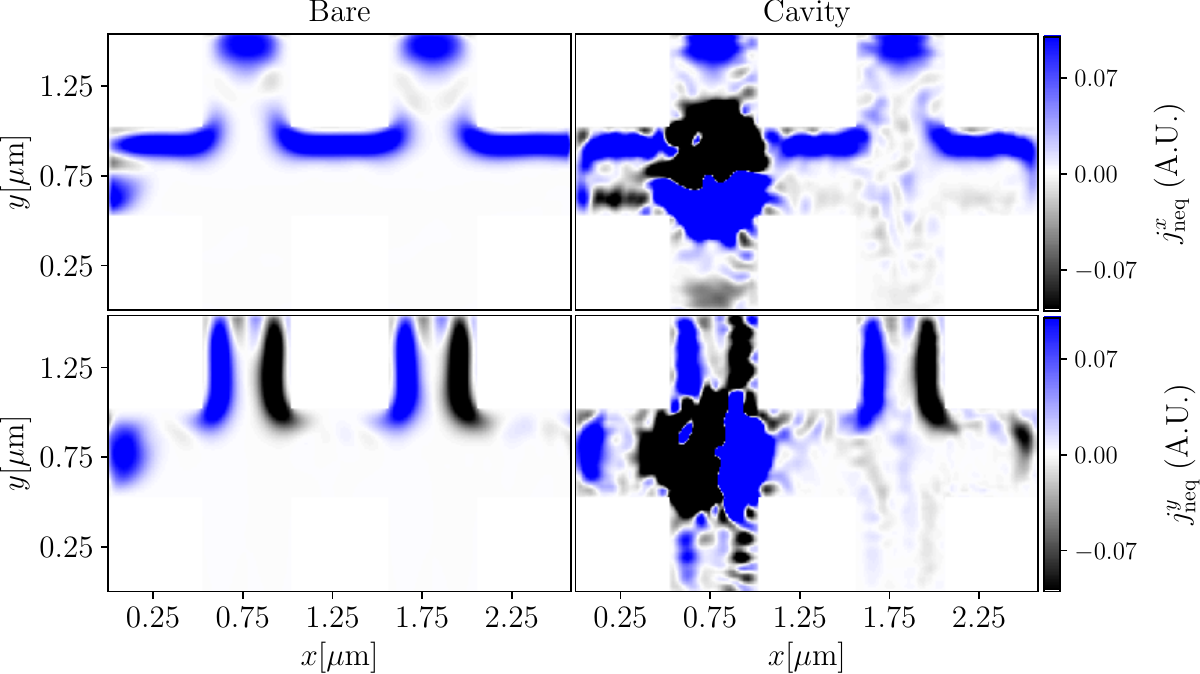}
    \caption{
        Spatial-dependent current density profiles on the Hall bar,
        with same parameters as in Fig. \ref{fig:6term_res}.
        There is a net current imposed between leads 1 (source, left)
        and 6 (drain, right).
        The Fermi level is set to $E_F = 1.075 \times \hbar \omega_{\mathrm{cyc}}$
        (corresponding to the vertical line in the plots showing the magnetoresistances).
        The first column corresponds to the bare currents, 
        while the second to the cavity-modified currents, showing inter-edge scattering.
        The first line represents the $x-$projection of the current density vector, 
        while the second line, the $y-$projection.
    }
    \label{fig:6term_currs}
\end{figure*}

In our study,
we have considered a six-terminal configuration
where the Hall bar is connected to six electrodes, 
with two of them serving as the source and drain. 
Fig. \ref{fig:DisProfAndModProf} illustrates the shape 
and dimensions of the sample used 
in our numerical simulations. We label the leads from 1 to 6, 
as in  Fig. \ref{fig:DisProfAndModProf}(a).
We calculate the $5\times5$ 
reduced resistance matrix
$
(\mathcal{R}_{pq})_{p,q \in \{1,...,5\}}
$
of the Hall bar numerically,
using the NEGF formalism
at any given $E_F$.
In quantum Hall experiments, 
the Hall and longitudinal resistances
are obtained by driving a current
through the source and drain 
and measuring voltage drops 
across zero-current contacts
on the same side of the current flow 
(longitudinal voltage $V_{\mathrm{L}}$)
and opposite sides (Hall voltage $V_{\mathrm{H}}$).
We can read out these voltages from the elements 
of the reduced resistance matrix $\mathcal{R}$.
The sign of the currents is defined as positive 
for incoming currents and negative for outgoing currents.
The current entering the source (lead 1)
is $I_1 = I > 0$.
Since there is only one drain (lead 6),
 we must have 
$I_2 = I_3 = I_4 = I_5 = 0$ 
and the conservation of current imposes $I_1 = - I_6 = - I$.
The Hall and longitudinal resistances 
are hence obtained as follows:
\begin{align}
    \mathcal{R}_{\mathrm{H}}
    &=
        \frac{V_{\mathrm{H}}}{I}
    =
    \left[
        \frac{V_4 - V_5}{I_1}
    \right]
    = \mathcal{R}_{41} - \mathcal{R}_{51}
    \, ,
    \\
    \mathcal{R}_{\mathrm{L}}
    &=
        \frac{V_{\mathrm{L}}}{I}
    =
    \left[
        \frac{V_3 - V_5}{I_1}
    \right]
    = \mathcal{R}_{31} - \mathcal{R}_{51}
    \, ,
\end{align}
where we have used Eq. \eqref{eqn:VIR}
to obtain the terminal voltages, 
for the given terminal currents $I_i$
and resistance matrix $\mathcal{R}$.

Let us now include the spin degree of freedom.
The Zeeman energy due to the magnetic field lifts the spin energy degeneracy.
The single body Hamiltonian $\hat{h}$
of the spinless electron given in 
Eq. \eqref{spinlessHam} 
is promoted to a spinor Hamiltonian
\begin{equation}
    \hat{h}_{\mathrm{spinful}}
    =
    \begin{pmatrix}
        \hat{h}_{\uparrow} & 0 \\
        0 & \hat{h}_{\downarrow}
    \end{pmatrix} \, ,
\end{equation}
with 
$
\hat{h}_{\uparrow} 
=
\hat{h} - \left( \frac{1}{2} g_{\star}\mu_B B \right) \hat{\mathbbm{1}}
$
and
$
\hat{h}_{\downarrow} 
=
\hat{h} + \left( \frac{1}{2} g_{\star}\mu_B B \right) \hat{\mathbbm{1}}
$. 
Here, $g_{\star}$ is the effective 
$g-$factor and $\mu_B$ the Bohr magneton.
In the absence of spin-orbit interaction and magnetic impurities, 
the spin channels are independent and conductances are simply additive, namely
$\mathrm{G}_{\mathrm{tot}} = \mathrm{G}_{\uparrow} + \mathrm{G}_{\downarrow}$.
The conductance can be expressed 
in terms of the spinless conductance matrix as:
\begin{equation}
    \mathrm{G}^{\mathrm{tot}}(E_F) 
    =
    \mathrm{G}\left(E_F + \frac{1}{2} E_Z \right)
    +
    \mathrm{G}\left(E_F - \frac{1}{2} E_Z \right) \, ,
\end{equation}
with $E_Z = g_{\star}\mu_B B $
and $\mathrm{G}$ being the 
conductance of the spinless model.
Since we have a quantum light-matter interaction conserving the electron spin, 
this relation also holds for the cavity-modified conductance matrix.
This conductance matrix is then used to compute 
the Hall and longitudinal resistances as usual.
Following similar arguments, 
it can be shown that the following holds 
for the local currents and density of states
\begin{align}
    \nonumber
    \bm{j}_{\mathrm{neq}}^{\mathrm{tot}}
    (\bm{r}_k; E_F)
    &=
    \frac{e^2}{h}\sum_{q=1}^{N_{\mathrm{L}}}
    V_q
    \sum_{j}
    \frac{
        \bm{\bm{r}_j - \bm{r}_k}
    }{
        a^2
    } \times \\
    & \times 
    \Im{ \Bigg\{
        \tilde{t}_{kj} 
        \sum_{\sigma}
        G^{n;q}_{jk} \left(E_F + \sigma \frac{1}{2} E_Z \right)
        \Bigg\}
    } \, ,
    \label{eqn:current6termRaw}
    \\
    \nonumber
    \rho^{\mathrm{tot}}(\bm{r}_k ; E_F) 
    &=
    \rho \left( \bm{r}_k ; E_F + \frac{1}{2} E_Z \right) + \\
    & \quad \quad \quad\quad\quad\quad
    +
    \rho \left( \bm{r}_k ; E_F -   \frac{1}{2}  E_Z \right) \, .
\end{align}
To compute the local current density
in the six-terminal device, 
we use the resistance matrix $\mathcal{R}$ 
to express the voltages, 
leaving only the imposed current $I$ 
as a free parameter.
We replace the values of $V_q$ as follows:
we can set $\mu_6 = \sum_{0\leq q \leq 5} \mu_q / 5 = E_F$, 
to have $V_6 = 0$ in turn as discussed earlier.  
By imposing our usual terminal currents 
and using the current voltage relation, we get:
\begin{align}
    \nonumber
    \bm{j}_{\mathrm{neq}}(\bm{r}_k)
    &=
    \frac{I}{a}
    \sum_{j}
    \left(
    \frac{
        \bm{r}_j - \bm{r}_k
    }{
        a
    }\right)
    \times \\
    &
    \times
    \left[
        \sum_{q=1}^{5}
        \left(
            \frac{e^2}{h}
            \mathcal{R}_{q1}
        \right)
        \sum_{\sigma}
        \Im{ \left\{
            t_{kj}
            G^{\mathrm{n};q}_{jk}
            \left(
            E_F + \frac{\sigma}{2} E_Z
            \right)
        \right\} }
    \right] \, .
\end{align}
The resistance matrix elements 
$\mathcal{R}_{q1}$ are computed
by inverting the two-spin 
channel conduction matrix.

We have computed the Hall and longitudinal resistances
for a range of Fermi levels and have reported our results
in Fig. \ref{fig:6term_res}.
Electronic disorder has been included in our sample, 
as shown in Fig. \ref{fig:DisProfAndModProf}(c).
This potential is generated 
as a random function with a finite correlation length. 
The cavity mode we consider 
has a spatially varying profile, 
depicted in  Fig. \ref{fig:DisProfAndModProf}(b).
We have attempted to mimic the strong gradients along the 
Hall bar edges, reported in \cite{Enkner}.
For more details on the 
cavity parameters refer to the caption in 
Fig. \ref{fig:6term_res}. We emphasize that with our theoretical framework we can consider electronic disordered potentials and cavity mode profiles of arbitrary shape, so what considered is just an illustrative and representative example.

The bare (no-cavity) Hall resistance 
shows distinct steps at quantized values
$
\mathcal{R}_{\mathrm{H}}
=
h / (\nu e^2)
$,
where $\nu$ represents the number 
of fully filled Landau levels (including spin). 
To be precise, 
$
\nu (E_F)
=
\sum_{n \sigma} 
\mathbf{1}_{
    E_{n\sigma} < E_F
} 
$,
where $E_{n\sigma} = E_{\mathrm{cyc}}(n+1/2) + \sigma E_Z/2$,
$n \in \mathbb{N}$ and $\sigma \in \{+1, -1\}$.
The longitudinal resistance 
$\mathcal{R}_{\mathrm{L}}$ vanishes 
whenever $E_F$ lies between two consecutive $E_{n\sigma}$ values
and exhibits peaks when $\mathcal{R}_{\mathrm{H}}$ 
steps into a new plateau.
These transition Fermi energies are
$E_F \in \{ E_{n \sigma} \}_{n \sigma}$.
This magnetoresistance phenomenology is typical of the well-known 
integer quantum Hall effect.
It occurs because the electrons 
flow along chiral edge modes without dissipation, 
whlie the bulk acts as an insulator.
The only way to disrupt this effect 
is by enabling electrons to scatter `backwards' \cite{Faist1991_InterChannelScattering}, 
which means traveling to the edge 
that flows in the opposite direction.
This process is called backscattering and is absent 
in the quantum Hall regime due to the insulating bulk.
The edge states are said to be topologically protected.
It can happen in narrow quantum Hall bars, but if the width is large enough inter-edge scattering becomes negligible.

In sharp contrast, for the considered Hall bar width, the cavity Hall resistance quantization breaks down and correspondingly a finite longitudinal resistance emerges at integer filling factors. 
The odd numbered plateaus (those for which $\nu$ is odd)
are more sensitive to the 
cavity-induced long-range hopping if the Zeeman energy splitting is smaller than the cyclotron energy.
This is because, since $E_Z \ll \hbar \omega_{\mathrm{cyc}}$,
at odd integer filling the plateaus are closer in energy.
These findings are in agreement with 
experimental evidence in \cite{Appugliese2022}.
In Fig. \ref{fig:6term_currs},
we show the non-equilibrium current density
when current flows from terminal 1 to 6 
at $E_F = 1.075 \times E_{\mathrm{cyc}}$.
The bare currents flow exclusively 
along the edge without any resistance. 
If we look at the $x-$projection of the currents, 
they are all in blue, meaning positive, 
flowing from left to right.
In presence of the cavity, 
there are currents flowing backwards,
carrying charge to the opposite edge
and thus increasing resistance to the 
imposed current that flows from left to right. This is a result of cavity-mediated inter-edge scattering \cite{Ciuti2021}.

To conclude this section, we would like to make a few remarks about the computational complexity of our cavity-modified quantum transport calculations. 
The continuum limit of the discretized model
is obtained when 
$E_F \ll 2t = \hbar^2/(m_{\star} a^2)$, 
with $a$ being the lattice spacing.
Here we have scanned $E_F$ over a range 
$ \{ \nu \hbar \omega_{\mathrm{cyc}} ; \nu \in [1,3]\}$.
The continuum limit condition for $\nu$ is hence given by
$ \nu \ll (\ell_{\mathrm{cyc}} / a)^2$.
For a lead width 
$l_{\mathrm{W}} = 0.5 \, \mathrm{\mu m}$,
we can express the continuum limit condition again as 
$\nu \ll 0.0025 \times (N_{\mathrm{W}}^2/B[\mathrm{T}])$,
with $N_{\mathrm{W}}$ being the number of discretisation points 
along the lead's width and 
$B[\mathrm{T}]$ being the magnetic field in units of Tesla.
The reported simulations have been carried out 
for $N_{\mathrm{W}} = 25$ and $B = 0.1 \, \mathrm{T}$.
In order to simulate a field $16$ times stronger, namely 
$B=1.6 \mathrm{T}$,
at the same convergence to the continuum limit rate, 
we would need a grid $4$ times finer \footnote{ 
The dimension of the dense (not sparse) Hamiltonian matrix in our case is 
$ 7 \text{'} 000 \times  7 \text{'} 000 $, which is close to the memory limit in our computer cluster}.

\begin{figure}[t!]
    \centering
    \includegraphics[scale=0.5]{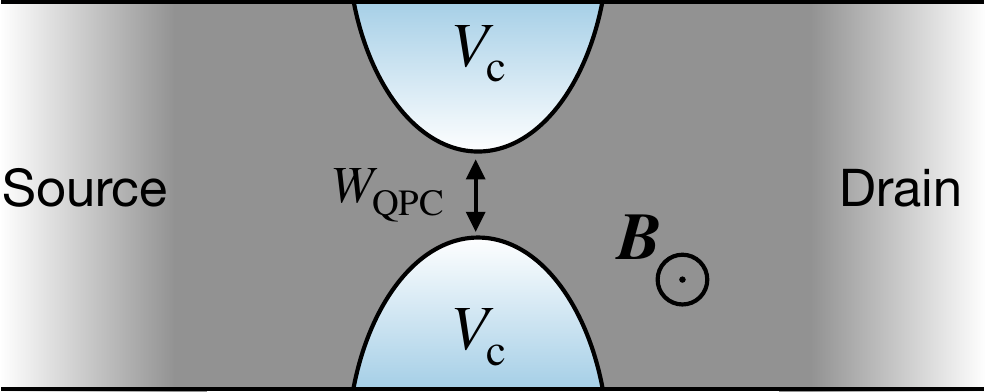}
    \caption{
        Sketch of a Quantum Point Contact (QPC). 
        The cyan blobs are regions with a constant and high constriction 
        potential barrier $V_{\mathrm{c}} \gg \hbar \omega_{\mathrm{cyc}}$. 
        The distance between the boundaries of these regions is $W_{\mathrm{QPC}}$. 
        In the following calculations, 
        we have considered a barrier region defined by the boundary
        $
        f_{\mathrm{bound}}(\tilde{x}) 
        = 
        \frac{L_y - W_{\mathrm{QPC}}}{2}
        \exp
        \left[
            - \frac{(\tilde{x}/\delta)^2}{1-(\tilde{x}/\delta)^2}
        \right]
        $ 
        for 
        $-\delta \leq \tilde{x} \leq \delta$
        and 
        $f_{\mathrm{bound}}(\tilde{x}) = 0$ 
        otherwise. 
        In all the reported calculations, we have taken 
        $\delta = 20 \, \mathrm{nm}$. 
        The horizontal coordinate $\tilde{x}$ is zero in the middle of the barrier region.
       }
    \label{fig:QPC_sketch}
\end{figure}

\section{\label{sec:QPC}Quantum Point Contacts}
A quantum point contact (QPC) 
is a narrow constriction between 
two wider electron gas regions, 
often fabricated on a 2DEG system \cite{QPC_beenakker}.
We shall study the case when the 2DEG is coupled to
a strong perpendicular magnetic field.
Given the chiral nature of edge states 
in quantum Hall physics, 
bringing the edges close together 
generates interesting effects, 
as the states on opposing sides 
propagate in opposite directions. We consider a 2DEG confined to a rectangular slab 
under a strong perpendicular magnetic field. 
The Hamiltonian governing this system 
is the one used for the six-terminal Hall bar
(see Eq. \eqref{eqn:Ham2DEG}),
but the geometry here is simpler, 
consisting of a rectangular $L_x \times L_y$ slab connected 
to only two leads along the $x-$axis.
To impose a constriction in the transverse direction, 
we consider a strong and constant potential barrier, as shown in Fig. \ref{fig:QPC_sketch}. In the following, we will simplify our analysis
by considering only one spin channel.
We are interested in investigating how 
the width of the constriction, $W_{\mathrm{QPC}}$,
and cavity vacuum fields 
influence the two-terminal conductance,
the local density of states,
and the local current densities.

\begin{figure*}[t!]
    \centering
    \includegraphics[scale=0.47]{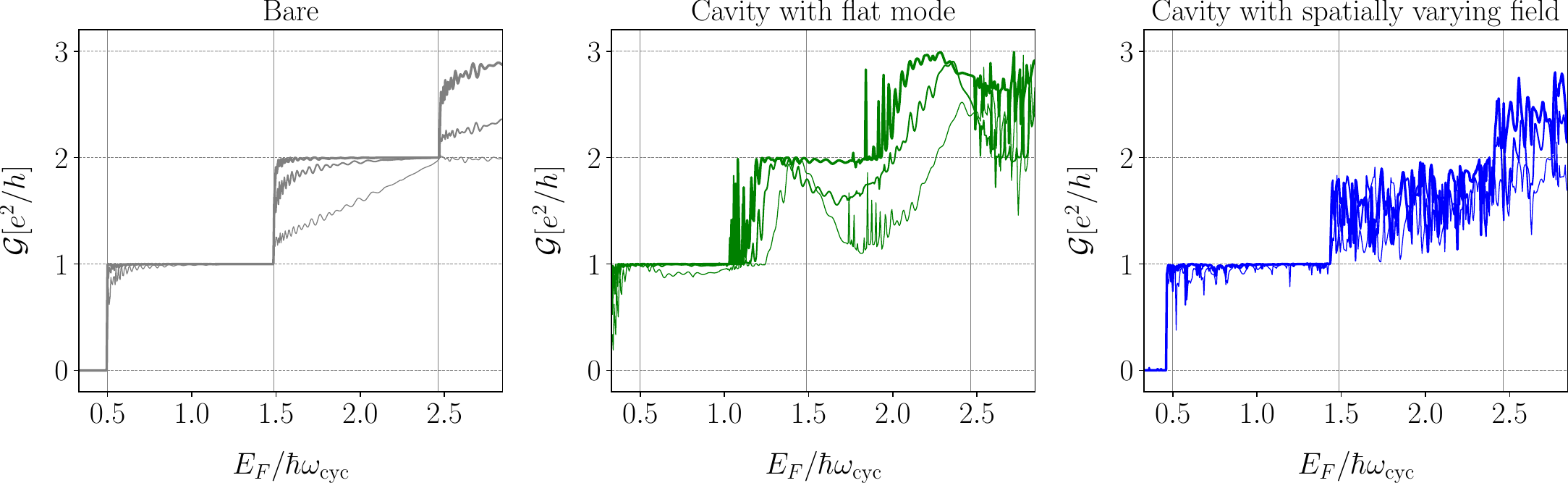}
    \caption{
        Two-terminal conductance (only a single spin channel is considered) of a quantum point contact 
        with varying constriction widths as a function of Fermi energy. 
        From left to right, the first plot shows the bare conductance (no cavity coupling),
        the second shows conductance for a sample with flat cavity mode 
        (no spatial gradients)
        and the third one for a spatially-varying mode.
        Parameters of calculations:
        rectangular slab with dimensions 
        $
        L_x \times L_y 
        =
        600 \; \mathrm{nm}
        \times
        150 \; \mathrm{nm}
        $, $m_{\star} = 0.067 \times m_e$, 
        $B = 6 \; \mathrm{T}$,   $a = 2 \; \mathrm{nm}$,  
        $V_{\mathrm{c}} = 100 \times \hbar \omega_{\mathrm{cyc}}$.
        The distance between the lower and upper regions, 
        $W_{\mathrm{QPC}}$,
        has been set to different values: 
        $32 \, \mathrm{nm}$,
        $40 \, \mathrm{nm}$
        and
        $48 \, \mathrm{nm}$ 
        and in the plots, the linewidths are proportional to the constriction width.
        Other parameters: 
        $\omega_{\mathrm{cav}} = 2\pi \times 10^{12} \mathrm{rad/s}$ 
        and
        $\eta=3\times 10^{-11}$ 
        (see caption of Fig. \ref{fig:6term_res} for details 
        on how the amplitude of the vector potential is defined).
        The polarisation vector is along the $y$ direction.
        For the third plot, the mode profile is $x-$dependent:
        $\mathcal{A}(x) = 0.5$ for 
        $0 \leq x[\mathrm{nm}] \leq 225$,
        $ = 1.5$ for 
        $375 \leq x[\mathrm{nm}] \leq 600 $,
        and increasing linearly from $x = 225 \mathrm{nm}$ to $x = 325 \mathrm{nm}$.
    }
    \label{fig:QPCs_flat_noFlat}
\end{figure*}

\begin{figure*}[t!]
    \centering
    \includegraphics[scale=0.6]{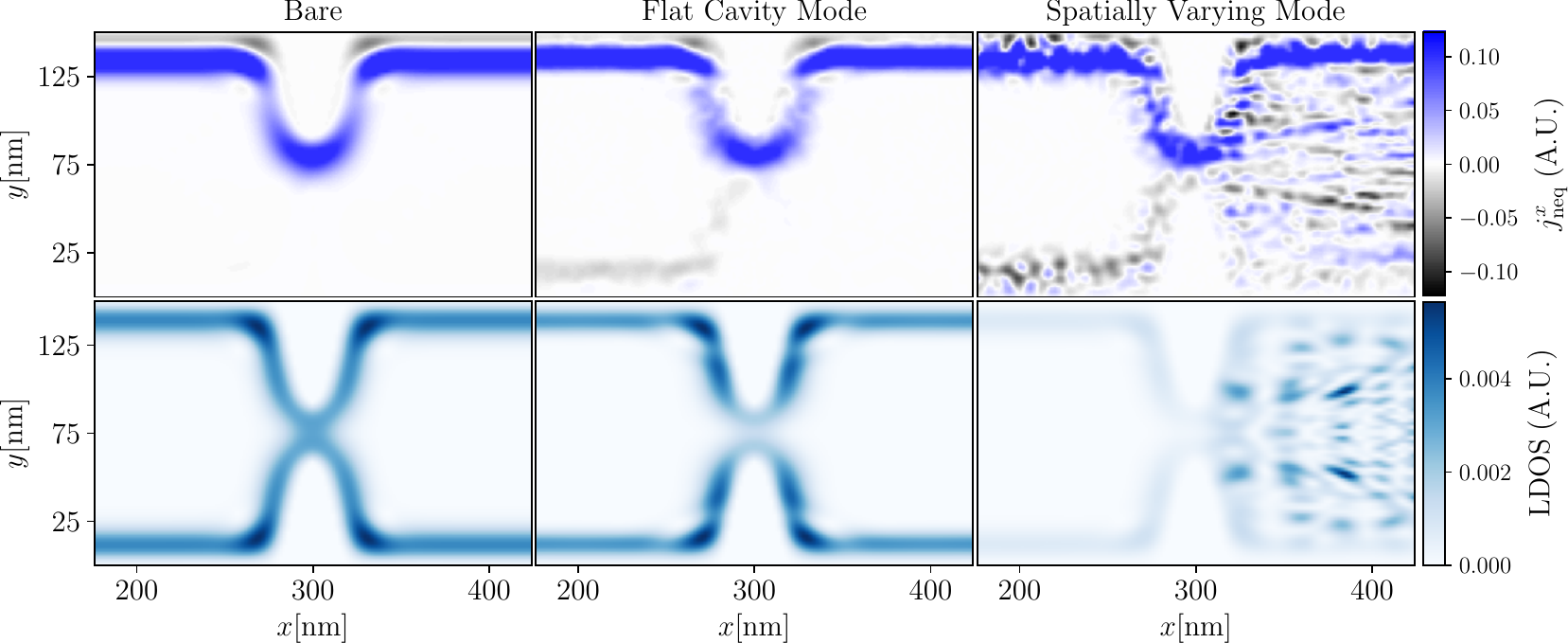}
    \caption{
        Non-equilibrium current density and local densities of states 
        for Hall bar quantum point contacts of width 
        $W_{\mathrm{QPC}} = 32 \, \mathrm{nm}$ 
        at filling $E_F = 0.825 \times \hbar \omega_{\mathrm{cyc}}$
        with no cavity, flat mode and a gradient field. 
        We have imposed a potential difference $V_L > V_R$, 
        so that current flows from left to right.
        The electronic system and the cavity parameters are the same as 
        the one described in Fig. \ref{fig:QPCs_flat_noFlat}.
        The first row of panels shows the $x-$projection of the current density $\bm{j}_{\mathrm{neq}} \cdot \bm{e}_x$. 
        The second row of panels shows the local density of states.
        Only the central region, close to the constriction is showed here.
    }
    \label{fig:3plotsBC}
\end{figure*}


In Fig. \ref{fig:QPCs_flat_noFlat}, 
first plot from the left, 
we present the conductance curve as a function 
of Fermi energy for three different values of $W_{\mathrm{QPC}}$, 
with thicker lines representing larger openings.
The first notable feature is the quantization of conductance, 
which is characteristic of quantum Hall systems. 
This conductance quantization persists in QPCs 
even without the need of a magnetic field.
As expected from numerous experimental
and theoretical works,
reducing $W_{\mathrm{QPC}}$ brings the edges closer
and introduces backscattering leading to an  
increased resistance.
We see that conductance is toned down
as the width becomes smaller. 
This phenomenon is more prominent at higher filling, 
since the edge states penetrate more in the bulk.
Bringing together two edge states from opposite sides 
allows electrons to flow in the opposite direction 
of the imposed voltage gradient.
This phenomenon disrupts the system's topology.

In the second and third plots of Fig. \ref{fig:QPCs_flat_noFlat}, 
we present the conductance curves modified by the cavity. 
While there is a slight shift in energy levels, 
the most notable feature is the near-complete destruction of quantized plateaus.
The second plot corresponds to a flat cavity mode, 
while the third reflects a cavity mode 
with a gradient at the contact. 
Further details are provided 
in the caption of Fig. \ref{fig:QPCs_flat_noFlat}. Despite the destruction of quantization, 
the conductance still exhibits the characteristic 
of increasing with larger openings. 
The spatially varying cavity mode 
appears to have a stronger impact on the system, 
due to a stronger breaking of translational invariance. 
This results in significant deviations from quantization, 
caused by cavity-mediated hopping \cite{Ciuti2021}.

In the first row of Fig. \ref{fig:3plotsBC}, 
we plot the $x-$component of the non-equilibrium currents
at a fixed Fermi level $E_F = 0.825  \times E_{\mathrm{cyc}}$, 
just above the first Landau level (where only one edge state is present),
and a fixed quantum point contact (QPC) opening
$W_{\mathrm{QPC}} = 32 \mathrm{nm}$
for the three cases: bare, flat cavity mode, 
and spatially varying cavity mode.
It is clear that cavity coupling creates backscattering
along the lower edge through cavity-mediated electron hopping. 
Moreover, it leads to the formation of new bulk states
in the case of a spatially varying cavity mode
as witnessed by the new peaks in the local density of states.

\section{\label{sec:AB}Aharanov-Bohm interferometers}
The last type of device we consider in this article
are electron interferometers based on the celebrated Aharonov-Bohm (AB) effect, which is remarkable as it shows that
charged particles are influenced by electromagnetic potentials, 
even in regions where the magnetic and electric fields are strictly zero \cite{Aharanov}. 
An AB interferometer can be achieved by splitting an electron beam and directing it around a region with a confined magnetic field, such as inside a solenoid. 
Despite the electrons never entering the magnetic field, 
they experience a phase shift due to the non-zero vector potential. 
This phase shift leads to observable interference patterns, 
such as conductance through an AB interferometer, 
which we shall study here.

\begin{figure}[t!]
    \centering
    \includegraphics[scale=0.34]{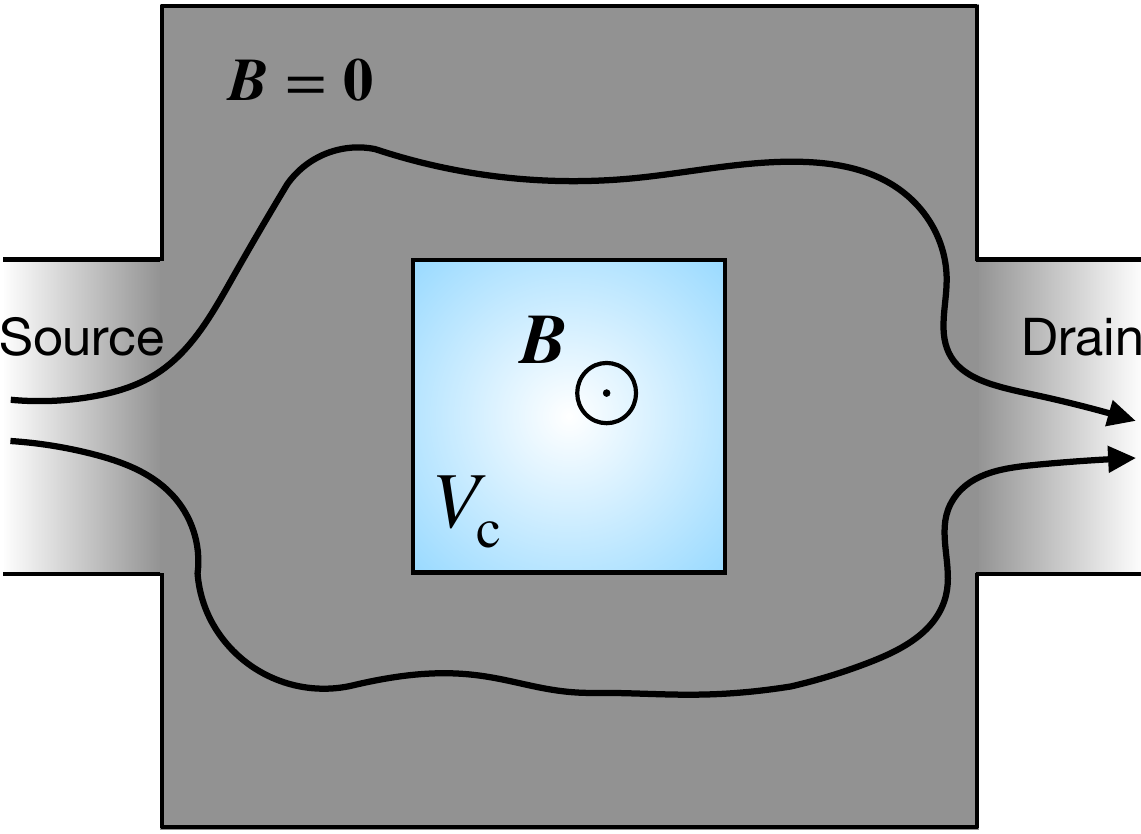}
    \caption{
        Sketch of the Aharanov Bohm interferometer. 
        The cyan square in the middle is pierced by a magnetic field $\bm{B}$. 
        We have imposed a very large contriction potential $V_{\mathrm{c}}$ potential on that part 
        to prevent electrons from flowing on the region with magnetic field.
        The two arrows represent two paths the electron could take in going from left to right. 
        The source and drain here serve also as contacts.
    }
    \label{fig:ABSketch}
\end{figure}

\begin{figure}[t!]
    \centering
    \includegraphics[scale=0.53]{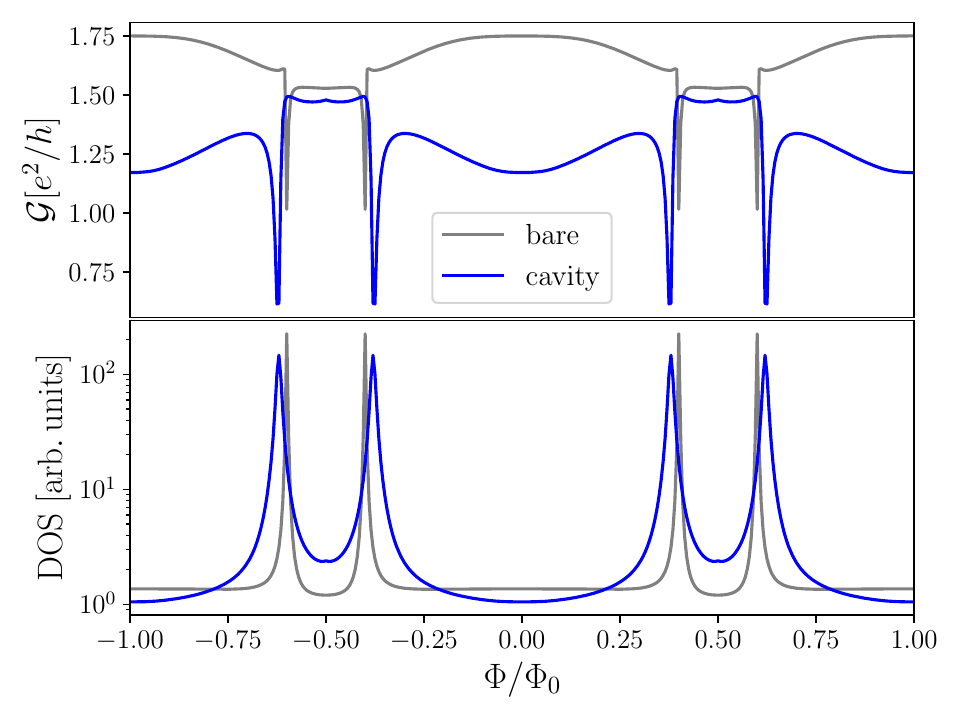}
    \caption{
        Conductance (upper panel) and density of states 
        (lower panel) as a function of magnetic flux over
        two periods.  The grey curve is for the bare case, while blue refers to the cavity case. The Fermi energy is $E_F = 10 \, \mathrm{meV}$.
        The sample has dimensions 
        $
        L_x \times L_y 
        =
        150 \; \mathrm{nm}
        \times
        150 \; \mathrm{nm}
        $, 
        with the square hole in the middle 
        having a side length of $l = 50 \; \mathrm{nm}$.
        The leads are connected to the left and right sides of the sample 
        and their width is a third of the sample's side length, i.e. $50 \; \mathrm{nm}$. There is no electronic disorder in the calculation and the 
        cavity is polarised along the transverse direction with
        $\omega_{\mathrm{cav}} = 2\pi \times 10^{12} \mathrm{rad/s}$ 
        and
        $\eta=3\times 10^{-11}$.
        The wall potential is $V_c = 10 \, \mathrm{eV} \gg E_F$, 
        ensuring the electrons to remain outside of the $B \neq 0$ area. Other parameters: 
        $m_{\star} = 0.067 m_e$,  $a = 2 \; \mathrm{nm}$.
    }
    \label{fig:ABConductanceDos}
\end{figure}

\begin{figure}
    \centering
    \includegraphics[scale=0.7]{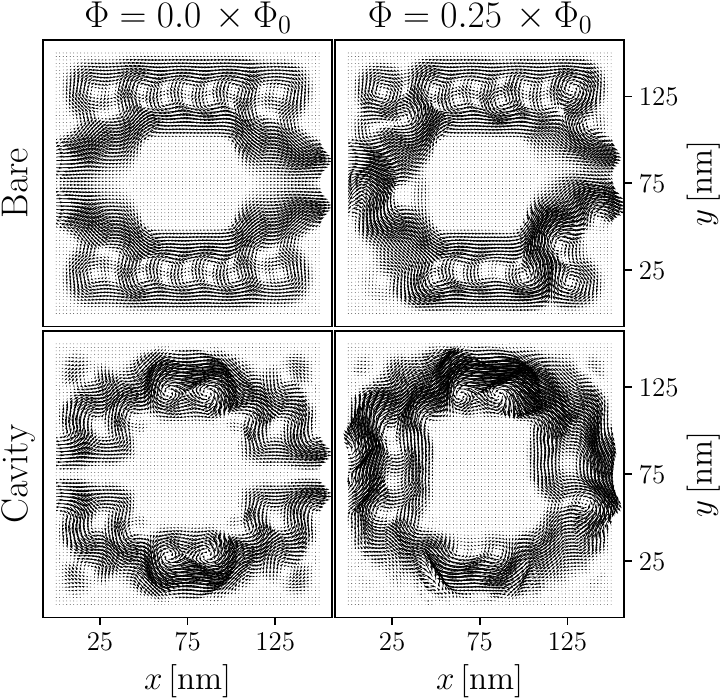}
    \caption{
        Vector field plot for the non-equilibrium current density $\bm{j}_{\mathrm{neq}}(\bm{r})$
        in the considered Aharanov-Bohm interferometer.
        The Fermi energy is set to $E_F = 12.9 \, \mathrm{meV}$.
        The flux is set to the values reported on top of each column.
        Other parameters are as in Fig. \ref{fig:ABConductanceDos}.
        }
        \label{fig:ABcurrents}
\end{figure}

The AB interferometer we consider 
consists of a 2D system with a hole in the middle, 
through which there is a magnetic flux going through.
The region where electrons can travel has zero magnetic field.  
In Fig. \ref{fig:ABSketch} we have drawn a sketch of our device. The phase picked by the electronic wavefunction around
the hole is proportional to the applied magnetic flux. For a given Fermi Energy $E_F$ the conductance is a periodic function of the applied magnetic flux, namely
$\mathcal{G}(E_F, \Phi + n \Phi_0) = \mathcal{G}(E_F,\Phi)$ 
for every integer $n$ and where $\Phi_0 = h/e$ is the flux quantum.
Note also  that, due to the Onsager-B\"uttiker relation \cite{OnsBut}, 
for any two-terminal device the conductance is an even function of the magnetic flux
$\mathcal{G}(E_F, \Phi) = \mathcal{G}(E_F, -\Phi)$.
Note that such properties remain unchanged in presence of the cavity coupling.
As an illustrative example, 
we have presented in Fig. \ref{fig:ABConductanceDos}
the variation of the conductance with respect to the applied magnetic flux, showing that such conductance is an even function of $\Phi$ and with period equal to $\Phi_0$. 


In Fig. \ref{fig:ABcurrents}, we report the current density for two different fluxes (left and right columns) for the case without the cavity (top row) and with the cavity (bottom row). This figure simply illustrates the sensitivity of the current patterns to the applied flux bias and how the cavity vacuum fields can non-trivially modify such pattern and hence the interference measured in the two-terminal conductance.

\begin{figure}
    \centering
    \includegraphics[scale=0.54]{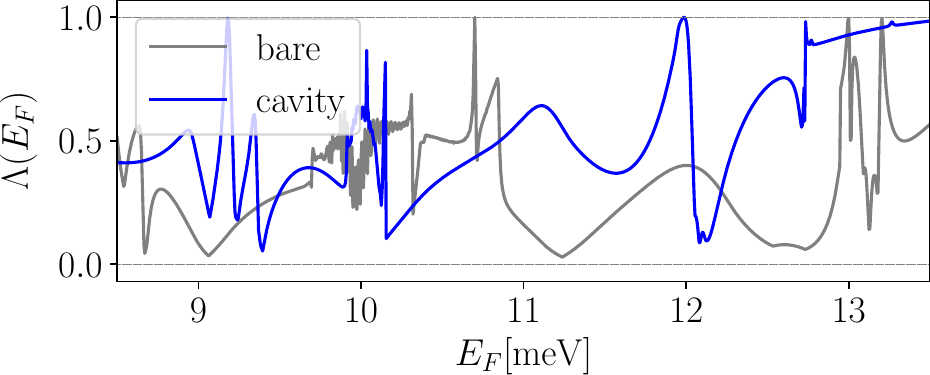}
    \caption{
        The visibility $\Lambda(E_F)$ of the considered Aharanov-Bohm interferometer.
        For each $E_F$, the conductance has been sampled 
        for 200 different equally-spaced values of $\Phi \in [0, \Phi_0]$.
        Other parameters are as in Fig. \ref{fig:ABConductanceDos}.
    }
    \label{fig:visibility}
\end{figure}

\begin{figure}
    \centering
    \includegraphics[scale=0.55]{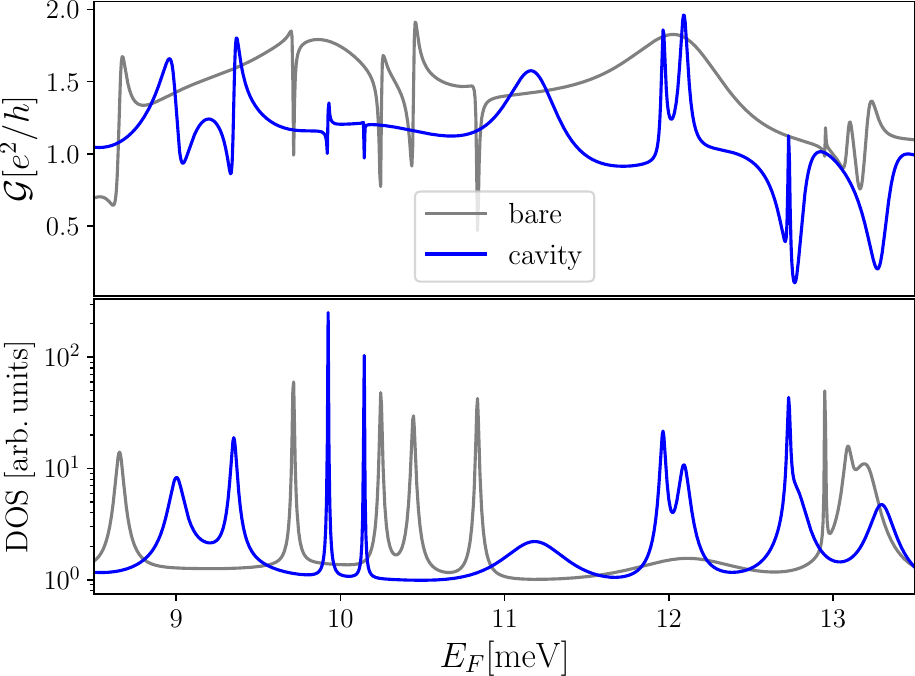}
    \caption{
        Conductance (upper panel) and density of states (lower panel)
        of the considered Aharanov Bohm interferometer as a function of Fermi energy. 
        The flux is fixed at $\Phi = 0.11 \times \Phi_0$. 
      Other parameters are as in Fig. \ref{fig:ABConductanceDos}.
    }
    \label{fig:condAB_fixedPhi}
\end{figure}

A key figure of merit of an AB devices is the interference visibility, which can be defined as:
\begin{equation}
    \Lambda(E_F)
    =
    \frac{
        \underset{\Phi}{\max} \, \mathcal{G}(E_F; \Phi)
        -
        \underset{\Phi}{\min} \, \mathcal{G}(E_F; \Phi)
    }{
        \underset{\Phi}{\max} \, \mathcal{G}(E_F; \Phi)
    } \, .
\end{equation}
In Fig. \ref{fig:visibility}, 
we have shown the visibility curve $\Lambda(E_F)$ 
in a relatively wide range of Fermi energies
for the bare device (grey line) and in presence of the cavity coupling (blue).
The irregular shape of the visibility is due to the complex density of states in the considered device geometry. This is illustrated in Fig. \ref{fig:condAB_fixedPhi}, which shows the conductance of the device in the same range of Fermi energies and the corresponding density of states. The influence of the cavity vacuum fields is both dramatic on the shape of the interference visibility, the conductance and the density of states. 

Note that density of states and conductance are correlated for both 
the bare and the cavity case.
A peak in the DOS curve corresponds to a kink in conductance.
From the DOS curve, we see that the cavity is able to alter the spectral 
properties of the material dictating new transport properties.
The very irregular structure of the curves we have reported here, 
comes as a consequence of the shape of the device.
In absence of magnetic field, 
the bulk conducts.
The said modes have irregular profiles 
due to the complicated shape of the boundary conditions.
Semi-classically, electrons can go in straight lines, 
bouncing and reflecting on the boundaries of the sample.
This billiard ball scheme couples many interference patterns.
We note that the cavity alters this structure by 
creating new paths for electrons to travel.

\section{Conclusions and Outlook}

In this work, we have explored the role of cavity-mediated electron hopping in influencing a variety of quantum transport devices, including multi-terminal quantum Hall bars, quantum point contacts, and Aharonov-Bohm electronic interferometers. Specifically, we have investigated the impact of cavity vacuum fields not only on macroscopic observables such as conductance and resistance but also on microscopic properties like current patterns and the local density of states. Notably, we have uncovered a rich phenomenology of cavity-mediated inter-edge scattering in narrow quantum Hall bars and quantum point contacts. Additionally, we have demonstrated that strong vacuum fields can qualitatively affect the interference visibility in Aharonov-Bohm devices. Our findings emphasize that strong vacuum fields—whether intentionally introduced or unintentionally generated by metallic contacts and gates acting as antennas—must be carefully considered in the design and analysis of mesoscopic quantum devices.

Looking forward, the formalism and methodologies developed in this study offer promising avenues for further exploration. One such direction is the application of this cavity-modified quantum transport framework to materials such as graphene and other tight-binding systems. Graphene, with its unique electronic properties—including a linear dispersion relation and topological features—presents an ideal platform for probing new regimes of quantum transport under the influence of cavity fields. Similarly, this approach can be extended to other low-dimensional materials, such as transition metal dichalcogenides (TMDs) and engineered lattices, where light-matter interactions may enable novel means of controlling electronic properties.

Moreover, a critical direction for future work is extending this framework to regimes where cavity vacuum fields cannot be adiabatically eliminated and light-matter entanglement becomes a dominant factor. These regimes, where the quantum nature of the cavity field is inseparable from the electronic degrees of freedom, could reveal fundamentally new quantum transport phenomena. Understanding such entangled light-matter systems will be key to harnessing quantum correlations and exploring the interplay between photonic and electronic excitations.

Future investigations could also broaden the scope of this formalism to include interactions between cavity fields and superconducting systems, fractional quantum Hall states, and even non-equilibrium dynamics in driven systems. These efforts have the potential to uncover new quantum transport phenomena and pave the way for the development of innovative quantum devices that exploit cavity-induced effects.

\begin{acknowledgments}
We acknowledge financial support from the French agency ANR 
through the project CaVdW (ANR-21-CE30-0056-0) and from the Israeli Council for Higher Education - VATAT.
\end{acknowledgments}

\appendix

\section{\label{apx:leadGF}Self-energy of the leads in a 2DEG}

In this Appendix, we report details about the determination of the self-energies due to the contact leads. 
Let us consider first a finite-size 2D system coupled to a single lead. In particular, the system is assumed to have a finite size in the transverse $y$ direction 
and to be semi--infinite in the longitudinal $x$ direction. 
The Hamiltonian of the such bipartite system reads:
\begin{equation}
    \begin{pmatrix}
    \hat{h} & \hat{\tau} \\
    \hat{\tau}^{\dagger} & \hat{H}
    \end{pmatrix} \, ,
\end{equation}
where $\hat{h}$
is the Hamiltonian of the 
lead,
$\hat{H}$ is the Hamiltonian of the 
device system
and $\hat{\tau}$ is the coupling matrix 
of the system to the leads. The contact Hamiltonian 
is
\begin{equation}
    \hat{h} = 
    4t \sum_{\iota} 
    \vert \iota \rangle \langle \iota \vert
    -t\sum_{\langle \iota,\kappa \rangle} 
    \vert \iota \rangle \langle \kappa \vert \, ,
\end{equation}
where the boundary conditions are implicit.
The system-lead coupling is only along the interface, namely
\begin{equation}
    \langle \iota \vert \hat{\tau}_q \vert i\rangle
    =
    \begin{cases}
    -t 
    & \text{if }
    \vert \bm{r}_{\iota} - \bm{r}_{i} \vert = a\\
    0 & \text{otherwise} 
\end{cases} \, ,
\end{equation}
where $\iota$ is a site of the lead and $i$ is a site of the system.
The corresponding retarded self-energy for the system due to the lead is given by
$
\hat{\Sigma}^{\mathrm{r}}
=
\hat{\tau}^{\dagger} \hat{g}^{\mathrm{r}} \hat{\tau}
$.
This means that
$
\langle i \vert \Sigma^{\mathrm{r}} \vert j \rangle
=
0
$
if either $i$ or $j$ is not along the system-lead interface and
$
\langle i \vert \Sigma^{\mathrm{r}} \vert j \rangle
=
t^2 \langle \iota \vert g^{\mathrm{r}} \vert \kappa \rangle
$,
where here $\iota$ (resp. $\kappa$) is the nearest neighbor 
site in the lead of the system site $i$ (resp. $j$).
This implies that we only need to compute $g^{\mathrm{r}}$
along the interface with the conductor. 

The lead Green's function is obtained by first
inverting the $(E - \hat{h} + \img \eta)$
matrix for a finite-size lead 
and then letting the number of sites $N_x$  along
its longitudinal dimension to infinity, 
keeping the transverse one fixed.
Note that it is essential to keep $\eta > 0$
and letting it tend to $0^{+}$ only after taking the
$N_x \to  +\infty$ limit.
Each site on the lead is denoted by an index $\iota$,
positioned at $\bm{r}_{\iota} = (x_{\iota},y_{\iota})$.
It is convenient to represent $\hat{h}$
in terms of its eigenstates $\vert \chi_{\alpha,\beta} \rangle $ defined by the eigenvalue equation
$
\hat{h} \vert \chi_{\alpha,\beta} \rangle
=
(
\varepsilon^{x}_{\alpha} 
+
\varepsilon^{y}_{\beta} 
)
\vert \chi_{\alpha,\beta} \rangle
$, with
\begin{align}
    \langle \iota  \vert  \chi_{\alpha,\beta} \rangle
    &=
    \chi^{x}_{\alpha}(x_{\iota}) 
    \chi^{y}_{\beta}(y_{\iota}) \, , \\ 
    \chi^{x}_{\alpha}(x_{\iota}) 
    &=
    \sqrt{\frac{2}{N_x+1}}
    \sin
    \left(
        \alpha \frac{\pi}{N_x + 1} \frac{x_{\iota}}{a}
    \right) \, , \\
    \varepsilon^{x}_{\alpha} 
    &=
    2t 
    \left[
        1 - 
        \cos
        \left(
            \alpha \frac{\pi}{N_x + 1}
        \right)
    \right] \, , \\
    \chi^{y}_{\beta}(y_{\iota}) 
    &=
    \sqrt{\frac{2}{N_y+1}}
    \sin
    \left(
        \beta \frac{\pi}{N_y + 1} \frac{y_{\iota}}{a}
    \right) \, , \\
    \varepsilon^{y}_{\beta} 
    &=
    2t 
    \left[
        1 - 
        \cos
        \left(
            \beta \frac{\pi}{N_y + 1}
        \right)
    \right] \, .
\end{align}
Note that here $\alpha \in \{1,...,N_x\}$
and $\beta \in \{1,...,N_y\}$ correspond to the longitudinal and transverse direction respectively. 
Using the spectral representation, 
we obtain the Green's function for the 
isolated lead on the surface, 
i.e. for the sites closest to the system 
($\bm{r}_{\iota}$ such that $x_{\iota} = a$, 
supposing without loss of generality 
that the system is situated at $x \leq 0$). Hence, we find
\begin{equation}
    \langle \iota \vert 
    \hat{g}^{\mathrm{r}}(E)
    \vert \kappa \rangle
    =
    \sum_{\mu = 1}^{N_y}
    \chi_{\mu}^{y}(y_\iota)
    \chi_{\mu}^{y}(y_\kappa)^{\star}
    \sum_{\nu=1}^{N_x}
    \frac{
        \chi_{\nu}^{x}(a)
        \chi_{\nu}^{x}(a)^{\star}
    }{
        E - \varepsilon^{y}_{\mu} - \varepsilon^{x}_{\nu} + \img \eta
    } \, .
\end{equation}
From this,  we can evaluate the Green's function of the 
infinite system in the limit $N_x \to +\infty$
by recognizing a Riemann sum
    \begin{align}
    \nonumber
    \langle \iota \vert 
    g^{\mathrm{r}}(E)
    \vert \kappa \rangle
    &=
    -\frac{1}{t}
    \sum_{\mu = 1}^{N_y}
    \chi_{\mu}^{y}(y_{\iota})
    \chi_{\mu}^{y}(y_{\kappa})^{\star} \times \\
    \nonumber
    &\times \lim_{N_x \to \infty}
    \sum_{\nu=1}^{N_x}
    \left( \frac{\pi}{N_x +1} \right)
    f_{\mu, E} \left( \nu \frac{\pi}{N_{x} + 1} \right)
    \, \\
    &=
    -\frac{1}{t}
    \sum_{\mu = 1}^{N_y}
    \chi_{\mu}^{y}(y_{\iota})
    \chi_{\mu}^{y}(y_{\kappa})^{\star}
    \int_{0}^{\pi}
    dx\,
    f_{\mu, E} \left( x \right)
    \, ,
\end{align}
where we have defined:
\begin{equation}
    f_{\mu, E} \left( x \right)
    =
    \frac{2}{\pi}
    \frac{ -\sin^2x }{
        \tilde{E}_{y \mu} + \img \eta + 2 \cos x 
    }
\end{equation}
and
$
\tilde{E}_{y \mu} = \frac{1}{t} (E - 2t - \varepsilon^{y}_{\mu})
$.
By extending the integral
over a full period and changing variables, 
it can be expressed as a contour integral 
over the unit circle:
\begin{equation}
    \int_0^{\pi} dx \, f_{\mu, E} \left( x \right)
    =
    \frac{1}{2 \img \pi}
    \oint_{ \mathbb{S}^{1} } dz \,
    \frac{z^2-1}{ 1 + (\tilde{E}_{y \mu} + \img \eta)z +z^2 } \, ,
\end{equation}
with $\eta \to 0^{+}$ in the end.
The result obtained using the residue theorem is the following:
\begin{widetext}
\begin{equation}
    \frac{1}{2 \img \pi}
    \oint_{ \mathbb{S}^{1} } dz \,
    \frac{z^2-1}{ 1 + (\tilde{E}_{y \mu} + \img \eta)z +z^2 }
    =
    \begin{cases}
        \dfrac{1}{2}
        \left(
            - \tilde{E}_{y \mu} + 
            \mathrm{sgn}(\tilde{E}_{y \mu} ) 
            \sqrt{\tilde{E}_{y \mu}^2-4} \,
        \right),
        &\text{  for  }
        \vert \tilde{E}_{y \mu} \vert > 2 \\ \\
        \dfrac{1}{2}
        \left(
            - \tilde{E}_{y \mu} + \img \sqrt{4 - \tilde{E}_{y \mu}^2} \,
        \right),
        &\text{  for  }
        \vert \tilde{E}_{y \mu} \vert \leq 2
    \end{cases} \, .
\end{equation} 
\end{widetext}
There is one crucial point we would like to point out. 
In case there is a lead connected to the scattering 
device subjected to a gauge field that picks 
the phase along the direction parallel 
to the transverse modes in the lead,
we should make sure that the wavefunctions
$\chi_{\mu}^{y}(y_j)$ account for that. 
Since there is no $B$ field in the leads, 
we must assure a constant gauge along the leads,
that extends by continuity from the device. 
This is handled by adding on the transverse 
modes the same phase as the one picked 
along the boundary of the device.
The expression of $\chi_{\mu}^{y}(y_j)$ is modified.
However, we obtain it numerically, 
by solving the 1D tight-binding particle 
on a finite chain,
with a gauge where a constant phase, 
equal to that on the boundary of the device, 
is picked along the chain.

\bibliography{./Bib_.bib}

\end{document}